\documentclass[12pt]{iopart}
\input{epsf.sty}
\usepackage{epsfig}
\usepackage{graphicx}
\usepackage{amsthm,amssymb,latexsym,amsfonts,times}
\usepackage{bm}
\usepackage{epstopdf} 

\begin{document}
\title{Observation of Magnetic Supercooling of the Transition to the Vortex State} 
\author{J.P. Davis$^1$, D. Vick$^2$, J.A.J. Burgess$^{1,2}$, D.C. Fortin$^1$, P. Li$^2$, V. Sauer$^{1,2}$, W.K. Hiebert$^2$ and M.R. Freeman$^{1,2}$}
\address{$^{1}$ Department of Physics, University of Alberta, Edmonton, Alberta, Canada T6G 2G7}
\address{$^{2}$ National Institute for Nanotechnology, Edmonton, Alberta, Canada T6G 2M9}

\date{Version \today}

\begin{abstract}  The magnetic hysteresis of an individual magnetic disk switching in and out of the vortex state has been exhaustively measured using nanomechanical torsional resonator torque magnetometry.  Each individual hysteresis loop pinpoints two sharp events,  a single vortex creation and an annihilation, with bias field precision of 0.02 kA/m. Statistical analysis of thousands of hysteresis loops reveals a dramatic difference in the sensitivity of the vortex creation and annihilation field distributions to the measurement conditions. The data sets measured at different magnetic field sweep rates demonstrate that the transition from the high-field state to the vortex state is not well modeled as a conventional thermal activation process in which it is assumed that the ``true" nucleation field is lower than any of the observed switching fields. Instead, the results are suggestive of the classic supercooling signature of a first-order phase transition, or more specifically here, its magnetic equivalent. This phenomenological evidence is consistent with a theoretical picture of the vortex nucleation process as a modified Landau first-order phase transition.  \end{abstract}

\maketitle

\vspace{11pt}
\section*{Introduction}

There are a wide variety of excellent techniques to study of nanomagnetism, but generally they all involve some form of averaging.  They either measure an array of disks \cite{Lau06}, average over multiple runs for adequate single to noise \cite{All03}, or are stroboscopic \cite{Fre98,Dem01,Zhi08}.  These types of measurements all have an implicit assumption.  The assumption underlying the interpretation of array results has been that each element has a well-defined switching distribution, narrower than that of the entire array, and the measured distribution from the array is inhomogeneously broadened because the individual elements are not identical. More insightful than bulk-averaged array measurements have been microscopy studies (MFM \cite{Zhu02}, Lorentz TEM \cite{Uhl05}) in which the switching events of each element are individually recorded during each hysteresis cycle.  This has led to the detection of unique behaviours within arrays, and even a correlation of coercivity with crystallite orientation for polycrystalline elements \cite{Lau08}.  One motivation for moving beyond these microscopy-based studies is that they are no longer a direct magnetometry, and therefore do not report precisely the same physical information as the traditional array experiments.  Micro-Hall bar \cite{Mih10} and micromechanical torque magnetometries \cite{Cha05,Dav10,Ban10} are well-suited among existing techniques for measuring actual M-H loops of individual nanomagnetic elements.  The Hall-based approach has yet to yield sufficiently rapid operation for the acquisition of statistically significant switching data sets from individual elements.  However increased sensitivity from scaling down the dimensions of micromechanical devices enables access to the regime where individual elements can be examined via torque magnetometry, in detail comparable to previous results from entire arrays.  The unambiguous result from this work is that individual polycrystalline elements themselves, in effect, manifest an entire ``array's worth" or distribution of hysteresis behavior, sensitively dependent on factors such as in-plane orientation of the applied magnetic field (on an angular scale comparable to the crystalline grain size divided by the lateral dimension of the element), and magnetic field sweep rate.  Most surprisingly, in measurements of the high-field state to vortex state transition of a single permalloy disk, field orientations are found for which an anomalously large downward shift of the vortex nucleation field distribution occurs with increasing sweep rate.  We present this as phenomenological evidence calling for further development of a recent theoretical description of this state change as a type of first-order phase transition \cite{Sav04}.  In contrast, no such behavior is observed for the reverse process, vortex annihilation, in which the sweep rate dependence is well modeled by a model for the thermal activation over an energy barrier \cite{Gar95}.

The micro/nano magnetic disk that we study in this paper has a large diameter-to-thickness ratio in zero applied magnetic field and its ground state is that of a magnetic vortex \cite{Gus01,Gus01b, Bad06}.  The magnetization curls in the plane of the disk and there is a vortex core associated with the singularity of curling magnetization, where the magnetization points out of the plane of the disk.  At large applied in-plane magnetic fields, the disk's energy is minimized by the parallel-spin state, in which the magnetization is aligned with the applied field.  At intermediate fields another magnetic state exists, the C-state, which can be thought of as a buckled form of the parallel-spin state.  The magnetization curls in the plane of the disk and a virtual vortex core lies outside of the disk \cite{Ant08}.  Other buckled states are possible, such as the S-state and onion state, but for simplicity we limit the discussion to the C-state since the exact details of the high-field state are not crucial for the physics that we discuss here, as we explain further below.

In the theory of Ref. \cite{Sav04}, vortex creation is described by a modified Landau first-order phase transition, where the applied field plays the role typically reserved for temperature \cite{Lan84} and there is a critical magnetic field, $H_c$, instead of a critical temperature.  As the magnetic field is lowered, $H_c$ is the highest magnetic field that results in vortex creation.  The order parameter associated with this phase transition is $\psi = s^{-1}$, where $s$ is the normalized distance between the center of the disk and the `rigid' vortex core \cite{Sav04}.  $\psi$ is zero in the parallel-spin state, describes a continuous transition to the C-state, and a discontinuous transition to the vortex state.  The free energy of the system can be expanded in even powers of $\psi$ and the transition from the parallel-spin state to the C-state is second-order, whereas the transition from either the parallel-spin state or the C-state to the vortex state is first-order.  The broken symmetry associated with the first-order phase transition is the out-of-plane magnetization of the vortex core.  The classic signature of a first-order phase transition is supercooling, associated with the nucleation of one state from the other \cite{Bin87}.  The amount of supercooling is dependent on the intrinsic energy barrier to nucleation, and extrinsic factors such as the surface roughness \cite{Sch95} and the rate at which the temperature (here the magnetic field) is lowered.  Experimental tests of this phase transition theory have not been undertaken, until now, because typical experiments on nanoscale magnetic elements involve averages over arrays of elements, or averages over many runs.  Ours does not.
\section*{Experiment and Results}

Our measurement is based on torque magnetometry, in which a magnetic element with magnetization $\boldsymbol M$, in the presence of an applied magnetic field, $\boldsymbol H$, experiences a torque, $\boldsymbol{\tau} = \boldsymbol{M} V \times \boldsymbol{H}$.  In the geometry shown in Fig.~1c, we are sensitive to the magnetization in the $\hat{x}$ direction, the torquing magnetic field is in $\hat{z}$ (out of the plane) and the torsion rod is oriented along $\hat{y}$.  The torque is detected as an out-of-plane displacement of the torsion paddle \cite{Dav10}.  We apply a constant AC magnetic field, ${H_z}$, at the resonance frequency of the torsional resonator.  The primary torsional mode, as shown in Fig.~1d, is at 2.81 MHz and has a mechanical $Q$ of 500.  We detect an optical interferometric signal through one of the large paddles at the resonant frequency, which is proportional to the out-of-plane displacement.  This is sensitive ($6 \times 10^{7} \mu_B$) to changes in the magnetization, ${M_x}$, of the magnetic element on the torsional resonator, which is varied by controlling the applied magnetic field along $\hat{x}$, ${H_x}$.  Further details can be found elsewhere \cite{Dav10}. 
	
\begin{figure}[t]
%%%%%%%%%%%%%%%%%   F I G U R E  1   %%%%%%%%%%%%%%%%%%
\centerline{\includegraphics[width=4.5in]{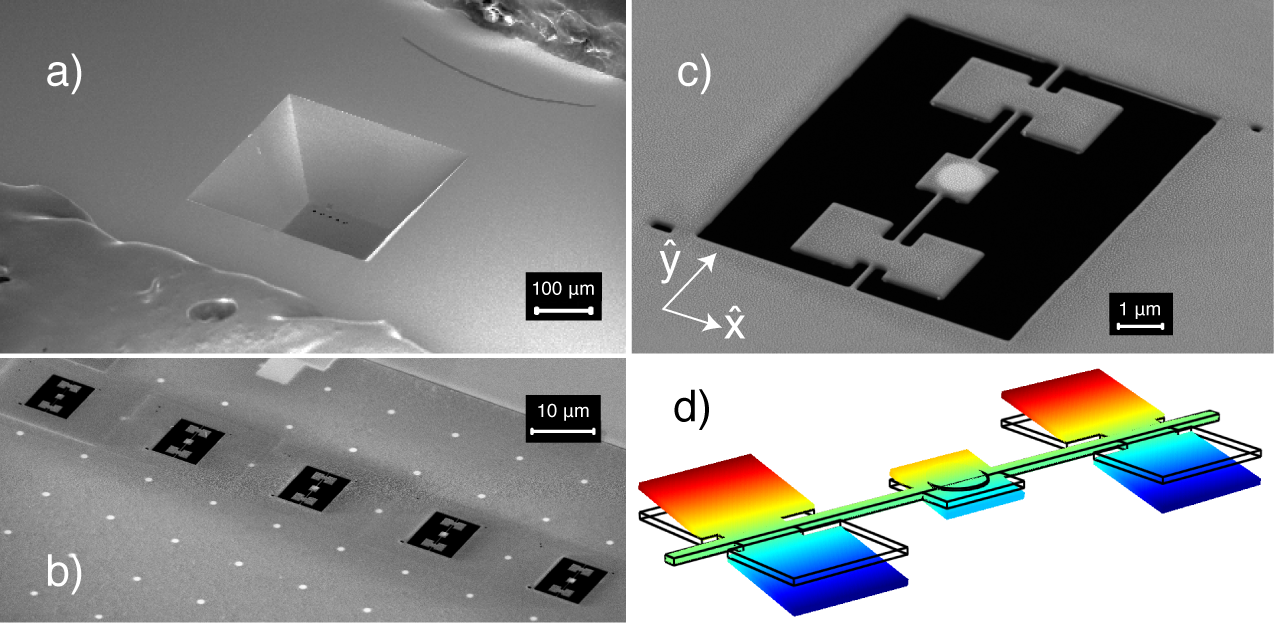}}
%%%%%%%%%%%%%%%%%%%%%%%%%%%%%%%%%%%%%%%%%%%%%
\caption{{\label{fig1}}  SEM micrographs of (a) the silicon nitride membrane on the silicon frame, (b) the array of magnetic disks visible through the silicon nitride with the nano-torsional resonators fabricated around the disks (just barely observable in a), and (c) the resonator used in this paper (also the lower right resonator in b).  (d) Finite element model of the primary torsional mode, in which all three paddles oscillate in phase.}
\end{figure}

Device fabrication starts with a commercially-available low-stress silicon nitride membrane on a silicon frame \cite{Nor}, Fig.~1a.  We deposit $42\pm2$ nm thick, as measured by calibrated atomic force microscopy, permalloy disks via collimated electron beam deposition at $10^{-9}$ mbar on the nitride using a stencil mask \cite{Des99} with $1~\mu$m diameter holes and $12~\mu$m spacing.  A $2$ nm gold anti-charging layer is sputtered onto the opposite side of the nitride membrane, and the frame is mounted, magnetic element side down, as in Fig.~1a.  The magnetic disks are visible through the silicon nitride with a scanning electron microscope, accurately aligned to be coincident with a focused-ion-beam mill.  This allows fabrication of a torsional resonator around a magnetic disk, Figs.~1b-c.  The resonator shown in Fig.~1c is used for all measurements in this paper.  After fabrication, the device is mounted $31~\mu$m from a silicon wafer, which acts as the back mirror of a low finesse Fabry-Perot cavity, and the entire structure is mounted directly onto a 2D Hall probe ($H_{x},H_{y}$) in a vacuum chamber \cite{Dav10}.  All measurements were performed at room temperature and the laser (HeNe, 632.8 nm, 3 $\mu$W hitting the resonator) for optical detection was kept at constant power for all measurements.  

\begin{figure}[t]
%%%%%%%%%%%%%%%%%   F I G U R E  2   %%%%%%%%%%%%%%%%%%
\centerline{\includegraphics[width=4in]{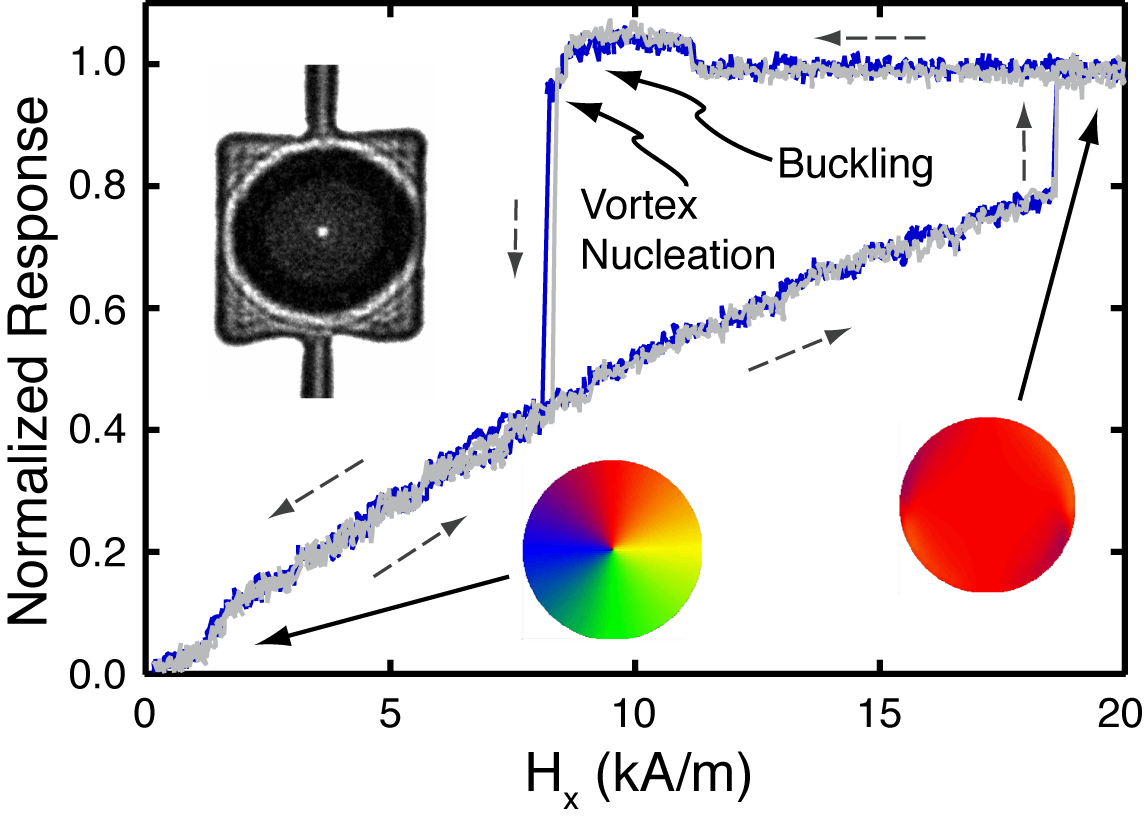}}
%%%%%%%%%%%%%%%%%%%%%%%%%%%%%%%%%%%%%%%%%%%%%
\caption{\label{fig2}  Interferometric response at 2.81 MHz versus applied bias magnetic field, $H_x$, resulting in magnetic hysteresis loops.  We show two such hysteresis loops, measured starting at saturation and following the dashed arrows, with different creation fields but identical annihilation fields.  Inset is a Lorentz micrograph \cite{Sch00} of the disk on the central paddle of the resonator used in these measurements.  The white spot in the center of the disk originates from the chirality of the vortex state, confirming the magnetic state of the exact disk used in the torque magnetometry.  We also show micromagnetic simulations of the in-plane magnetization.  The color map is: red is $\hat{x}$, blue is -$\hat{y}$, green is -$\hat{x}$, yellow is $\hat{y}$.}
\end{figure}

There are two applied magnetic fields in the experiment.  The first is the AC magnetic field at 2.81 MHz along $\hat{z}$.  Increasing this field increases the amplitude of the detected interferometric signal, but has no effect on the statistics of vortex creation and annihilation for our accessible field range.  Nonetheless, it is kept constant in all measurements (0.81 kA/m peak-to-peak).  The second magnetic field is the bias field which is applied along $\hat{x}$ and varies ${M_x}$ in the disk.  Sweeps of ${H_x}$ are shown in Fig.~2.  At large bias fields ${M_x}$ is saturated in the parallel-spin state.  This has been confirmed by micromagnetic simulations and is shown as the red disk in Fig.~2, although it can be seen that this is not a perfect parallel-spin state.  As the bias field is lowered there is a jump upwards then a small plateau, which we do not understand at this time, but does not affect in any way our conclusions.  Next there is a downward kink, and then a sharp drop as the magnetic vortex enters the disk and the magnetization drops abruptly.  This kink has been correlated with the buckling of the parallel-spin state to the C-state \cite{Rah03}.  The disk is relatively large at $1~\mu$m diameter so it is difficult to state with certainty that this is the parallel-spin to C-state transition, but regardless does not alter the interpretation of our results.  This is an important point, so we would like to reiterate it.  The exact nature of the high-field state is not relevant for our conclusions on the statistics of the vortex nucleation, as long as the high-field state is one with in-plane magnetization.  The reason for this is that the magnetic supercooling results from the breaking of a symmetry (the formation of the out-of-plane magnetization) upon vortex creation, which is identical for all transitions from states with entirely in-plane magnetization to the vortex state.  We can also be confident that our disk is not undergoing a double vortex transition.  This state has a very different remnant net magnetization \cite{Lau07a}, which would show up as a large non-zero signal at zero applied bias field.

At zero field the vortex sits near the center of the disk, which has been confirmed by Lorentz TEM microscopy \cite{Sch00} and micromagnetic simulations, shown as insets to Fig.~2.  The Lorentz TEM microscopy is unequivocal evidence that the remnant state of our disk is the single vortex state.  As the applied field is increased from zero the vortex is displaced from the center of the disk and is eventually annihilated, which corresponds to a smaller, but also abrupt change in ${M_x}$ \cite{Gus01,Gus01b}.  All of these features can be seen in Fig.~2.  The vortex creation and annihilation events occur over a small magnetic field interval, $\sim 0.02$ kA/m.  Experimental techniques that measure arrays of elements or average over multiple events would observe broadened transitions.

Fig.~2 displays two individual hysteresis sweeps with their corresponding vortex creation and annihilation fields.  Note that vortex creation occurs at two distinctly different magnetic fields, while the vortex annihilation fields are indistinguishable.  Repeated measurements of the events can be performed to learn about the statistics governing them.  In Fig.~3 we show a series of histograms showing the applied fields at which the vortex creation and annihilation events occur.  Figs.~3a and 3d correspond to low sweep rates, Figs.~3b and 3e to intermediate sweep rates, and Figs.~3c and 3f to high sweep rates.  The sweep rate is field dependent, since we sweep the position of a permanent magnet at a constant rate.

We observe that the vortex creation and annihilation field distributions are bimodal.  If during a field sweep down a vortex is created below 7 kA/m, then that same vortex annihilates during the subsequent field sweep up at a lower bias field than any vortex created above 7 kA/m.  We are not entirely confident of the origin of this phenomenon, but it is not sweep rate dependent and is not the focus of this manuscript.  We intend to explore this further in the future, but does not bear on our current conclusions.  Instead, we focus on the events ($93$\% of the total number of events we have measured) that correspond to vortex creation above 7 kA/m and the main annihilation peak.

It is seen in Fig.~3 that vortex creation occurs over a broader range of fields than vortex annihilation.  In the first-order phase transition model this spread in vortex creation fields occurs because the vortex is energetically favored to be inside the magnetic disk at the highest field at which creation occurs (8.5 kA/m).  But there is a barrier to nucleating the vortex core with out-of-plane magnetization.  This leads to supercooling of the high-field state and the spread of vortex creation fields.  It is the formation of the out-of plane magnetization from a high-field state that is entirely in-plane magnetized that requires this symmetry breaking and results in the supercooling.  No such symmetry breaking is required for vortex annihilation, since there exists in-plane magnetic moments in both the vortex state and the high-field state.  There is nonetheless a discontinuous transition, because of the finite amount of magnetization associated with the annihilation of the vortex core.

\begin{figure}[t]
%%%%%%%%%%%%%%%%%   F I G U R E  3   %%%%%%%%%%%%%%%%%%
\centerline{\includegraphics[width=5.25in]{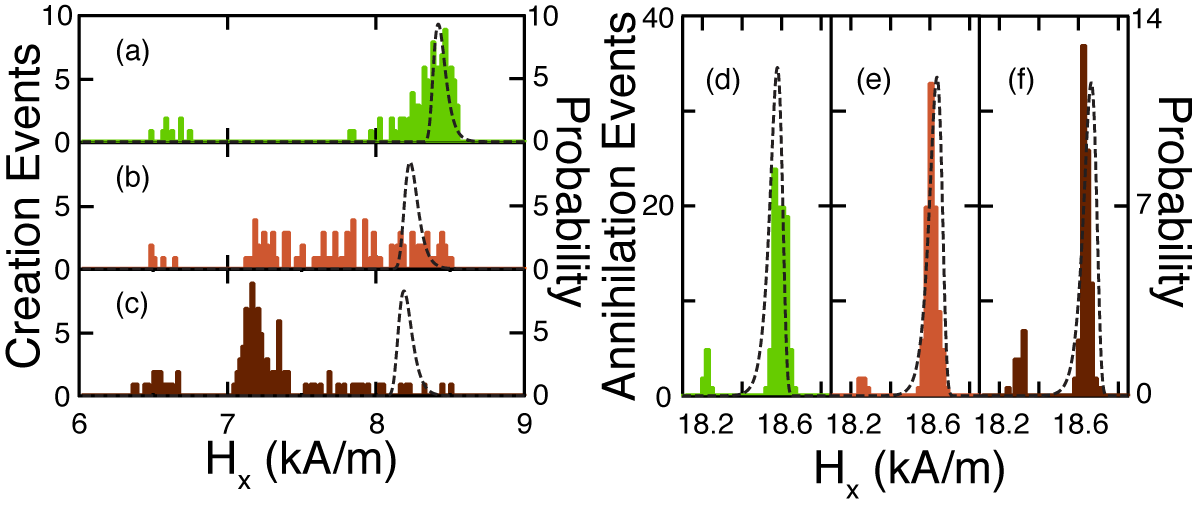}}
%%%%%%%%%%%%%%%%%%%%%%%%%%%%%%%%%%%%%%%%%%%%%
\caption{\label{fig3}  Histograms (left axis) of the magnetic fields of vortex creation and annihilation events binned by 0.02 kA/m.  Each panel has the same area and 100 events.  The dashed curve (right axis) is the probability distributions predicted by a thermally driven process over a field dependent energy barrier \cite{Gar95}, see text.  (a) Vortex creation at a sweep rate of 0.02 kA/m$\cdot$s, (b) 0.14 kA/m$\cdot$s, and (c) 0.36 kA/m$\cdot$s.  (d) Vortex annihilation at 0.07 kA/m$\cdot$s, (e) 0.43 kA/m$\cdot$s, and (f) 1.16 kA/m$\cdot$s.  The probability distributions match reasonably well to the bias fields, widths and sweep rate dependence for vortex annihilation, but do not for vortex creation.  The vortex annihilation is thermally activated but vortex creation involves different physics, magnetic supercooling.}
\end{figure}

Figs.~3a-c show that vortex creation occurs at lower bias fields for faster bias field sweep rates.  This rate dependence demonstrates that vortex nucleation occurs on laboratory time scales (the dynamics associated with the magnetization changes occur on picosecond time scales, but the vortex nucleation itself requires much longer waits).  The kink in Fig.~2 that we associate with the buckling transition from parallel-spin state to the C-state \cite{Rah03} always occurs at $8.55 \pm 0.06$ kA/m; it does not show supercooling or sweep rate dependence.  This is as expected from the theory of Ref.~\cite{Sav04}, in which this transition is of second-order and should not demonstrate supercooling, although confirmation of the second-order nature of the transition will require improvements in signal-to-noise.  In Fig.~4a, we plot the mean vortex creation field as a function of sweep rate.  Sweep rate dependence is a standard effect in supercooled first-order phase transitions \cite{Bin87, Sch95}.  Figs.~3d-f reveal that there is also a small amount of rate dependence to the vortex annihilation field distribution, which we plot in Fig.~4b.  The rate dependence of vortex creation is more than an order of magnitude larger than vortex annihilation.  

To analyze the sweep rate dependence, we compare our data with an Arrhenius relation for the probability per unit time of nucleating or annihilating a vortex, $dN/dt = f_0 e^{-E(H) / k_BT}$ \cite{Bro63, Sha99} with a field dependent energy barrier \cite{Gar95} and thermal activation over the barrier.  $f_0$ is an attempt frequency given by the order of magnitude of the frequency of spin waves at edge of the magnetic disk \cite{Par02}, $10^9$ Hz for vortex nucleation, and by the gyrotropic frequency for annihilation, $10^8$ Hz \cite{Iva04}.  $E(H)$ is the energy associated with the magnetic volume, $V$, involved in the switching event, given by $E = (1-H/H_a)^{3/2}\mu_0 M_s^2V / 2$ for annihilation and $E = (H/H_n -1)^{3/2}\mu_0 M_s^2V / 2$ for creation \cite{Gar95}.  $M_s$ is the saturation magnetization for permalloy ($800$ kA/m) and $\mu_0$ is $4\pi\times10^{-7}$ N/A$^2$.  There are three possible parameters in the model: $f_0$, $V$, and either $H_n$ or $H_a$ for nucleation or annihilation respectively.  We fix $f_0$ to be the values above (the fit is not very sensitive to $f_0$) and vary $V$ and $H_n$ ($H_a$) to match the bias field, width, and sweep rate dependence of the distributions.  Hence the parameters are overdetermined.

Probability distributions calculated from this model are shown as dashed curves in Fig.~3.   Vortex annihilation is well modeled with $H_a =19.4$ kA/m and a volume corresponding to a cylinder of 12.3 nm radius and 42 nm height.  This volume is slightly larger than expected, but reasonable, since the expected volume is the volume of the vortex core \cite{Gus01, Wac02}.   Our best fit to the creation data is shown in Figs.~3a-c, with $H_n =7$ kA/m and a cylinder of 4.3 nm radius and 42 nm height.  We note that in many thousands of runs we have never observed a vortex creation event above 8.5 kA/m.  This, along with the fact an $H_n$ greater than 7 kA/m is unphysical for this model with respect to the data, places additional constraints on the fit parameters.  The thermal nucleation model does not capture the physics of vortex creation, in contrast to the model's good fit to vortex annihilation.  Therefore, that a smaller volume is extracted for vortex creation than annihilation is unimportant.  Only the fit parameters, including the volume, for annihilation are considered relevant.  We conclude that the critical field for vortex creation is instead 8.5 kA/m with corresponding supercooling to lower fields.  The fundamental difference between vortex nucleation and annihilation is that there is a broken symmetry, the formation of an out of plane magnetization, that must occur to form the vortex state.  This is not the case for vortex annihilation because there is not such a symmetry breaking.  This transition is discontinuous since the annihilation of the vortex takes some finite out-of-plane magnetization associated with the vortex core and converts it to in-plane magnetization.  But there are already in-plane magnetic moments surrounding the vortex core, so there is no required symmetry breaking and no first-order phase transition.  Unfortunately, the modified Landau theory \cite{Sav04} has not yet been developed to compare probability distributions with our results as quantitatively as the thermal model.

\begin{figure}[t]
%%%%%%%%%%%%%%%%%   F I G U R E  4   %%%%%%%%%%%%%%%%%%
\centerline{\includegraphics[width=4.5in]{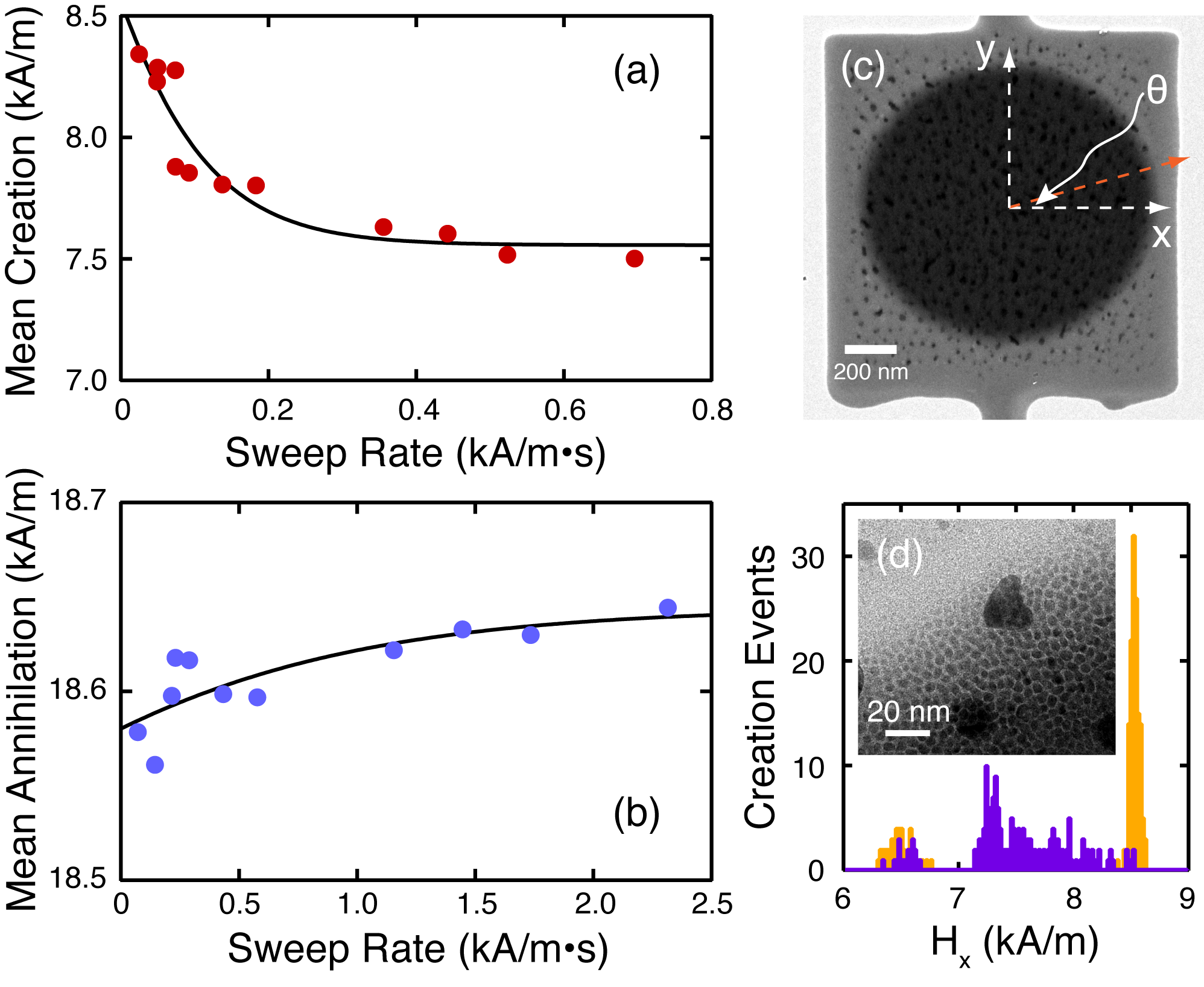}}
%%%%%%%%%%%%%%%%%%%%%%%%%%%%%%%%%%%%%%%%%%%%%
\caption{\label{fig4}  (a) Mean creation and (b) annihilation fields versus sweep rate.  The dashed curves show the sweep rate dependence of the mean of the probability distribution calculated from the Arrhenius relation.  We can fit the vortex annihilation data well over the full range of sweep rates.  For vortex creation this model does not capture the sweep rate dependence of the mean creation field.   In addition, the model does not fit the values for vortex creation at low sweep rates, since the shape of the creation data is not well fit as is seen in Figs.~3a-b.  (c) TEM image of the magnetic disk on the resonator, with the direction of bias-field rotation.  The dark flecks are the remnants of the gold on the opposite side of the nitride from the permalloy disk.  (d) Rotation of the bias field can eliminate the supercooling with all else the same (sweep rate 0.36 kA/m$\cdot$s): the dark (purple) histogram is 175 vortex creation events at 0 degrees and the light (orange) histogram is 175 events with the bias field rotated by 2 degrees.  The inset shows a TEM micrograph of the $\sim 5$ nm crystalline grains at the edge of the disk.  The large dark spots are the gold remnants.}
\end{figure}

Nonetheless, to further test the physical agreement with the theory \cite{Sav04} we rotate the applied bias field by small angles, $\theta$, shown in Fig.~4c.  Such rotation changes the location of vortex creation and annihilation along the magnetic disk's perimeter, which affects the amount of supercooling possible at a given temperature and sweep rate, due to the effective magnetic roughness of the disk's edge.  In Fig.~4d, we show two histograms.  The purple histogram shows significant supercooling and corresponds to 0 degrees, the same angle as all measurements discussed above.  The orange histogram is for the bias field direction rotated by 2 degrees in the plane of the resonator, and swept in magnitude while oriented in this new direction.  The supercooling effect is entirely removed and vortex creation becomes narrowly peaked, even at fast sweep rates.  Rotation back to 0 degrees consistently recovers the supercooling effect.  Moreover, the supercooling persists over angles between 0 and 2 degrees, and beyond 2 degrees, although the amount of supercooling is altered on the $\sim$ 0.5 degree scale.  

Two degrees corresponds to moving the vortex creation site along the disk circumference by $\sim 17$ nm, and 0.5 degrees by $\sim 4$ nm.  These numbers are consistent with the relevant length scales in the problem, which are the diameter of a vortex core, $\sim 10$ nm \cite{Gus01, Wac02}, and the permalloy crystalline grain roughness, $\sim 5$ nm, as measured by transmission electron microscopy (TEM), inset to Fig.~4d.  Note that there are no significant defects in the disk, as can be seen in a TEM micrograph of the disk (Fig.~4c).  It is by tuning the vortex entry position that we access the regime of sweep rate dependent supercooling on time scales that are experimentally obtainable.  One or more polycrystalline grains together act as a local magnetic roughness for vortex creation.  This has not been previously explored in measurements on single disks that should be sensitive to the local magnetic roughness \cite{Mih10}.  LLG simulations of magnetic disks show that notches and bumps on the perimeter of a disk, along the diameter perpendicular to the applied magnetic field direction, significantly affect the vortex creation field as compared to a disk with a smooth edge.  This confirms the crucial role of edge roughness in nucleating a vortex.  In essence we have access to a range of physical behaviors in a single magnetic disk.  We can tune from magnetically rough to magnetically smooth by small rotations of the applied field, which dramatically alters the transition from the high-field state to the vortex state.  In contrast, rotation of the applied field has little affect on vortex annihilation.  We have not observed any change in the width or sweep rate dependence of the probability distribution at any accessible angle.  We have observed small variations in the absolute value of the mean of the probability distribution, consistent with variations in the local magnetic energy for different polycrystalline grains.  

It is interesting to note that not only can one not measure an array of magnetic disks and assume that one is observing all of the physics in the system, but one cannot even measure a single location in a high quality magnetic disk and expect to observe all of the physics.  We believe that there is an entire realm of physical behavior that has been previously overlooked that is finally going to be accessible because of techniques that allow sensitive measurements of single nanomagnetic objects.

\section*{Conclusions}

In conclusion, we have observed a spreading and lowering of the vortex creation field in response to increasing magnetic field sweep rates.  This originates from supercooling of the high-field to vortex-state transition, due to a barrier to forming a vortex with out-of-plane magnetization.  This is phenomenological evidence that the transition to the vortex state in a single magnetic disk can be described in terms of a modified Landau first-order phase transition \cite{Sav04}.  The vortex supercooling in a polycrystalline disk can be eliminated by a small angle rotation of the applied bias field, which causes vortex creation and annihilation at a different position along the magnetic disk's perimeter with a different vortex nucleation barrier, associated with different local regions of magnetic crystalline grains.  These observations show previously unobserved physics in magnetic nanodisks \cite{Sav04}, and will affect their possible use as information storage media because of the long time scale for vortex core nucleation \cite{Sha99}.  Finally, we suggest that future measurements on magnetic disks with very low edge roughness and controlled localized defects for vortex creation and annihilation \cite{Lau07b,Par03}, will shed even more light on this topic.
\section*{Acknowledgments}

This work was supported by the Natural Sciences and Engineering Research Council, Canada; the Canadian Institute for Advanced Research; the informatics Circle of Research Excellence; Canada Research Chairs; and the National Institute for Nanotechnology.  Additionally, the Integrated Nanosystems Research Facility is supported through the Canada Foundation for Innovation and nanoAlberta.  We thank Don Mullin, Greg Popowich, Tony Walford and Steve Launspach for their contributions and we thank Kevin Beach for insightful discussions.

\end{document}